\documentclass[letter]{aa} 

\usepackage{amsmath}
\usepackage{graphicx}
\usepackage{graphics}
\usepackage{subfigure}
\usepackage{txfonts}
\usepackage{natbib}
\usepackage{xcolor}

\bibpunct{(}{)}{;}{a}{}{,}

\def\teff{\ifmmode T_{\rm eff} \else $T_{\mathrm{eff}}$\fi}

\def\ltsima{$\buildrel<\over\sim$}
\def\lsim{\lower.5ex\hbox{\ltsima}}

\newcommand{\hii}{H~{\sc ii}}
\newcommand{\ha}{\ifmmode {\rm H}\alpha \else H$\alpha$\fi}
\newcommand{\hb}{\ifmmode {\rm H}\beta \else H$\beta$\fi}
\newcommand{\lya}{\ifmmode {\rm Ly}\alpha \else Ly$\alpha$\fi}

\newcommand{\hep}{He$^+$}

\newcommand{\heii}{He~{\sc ii}}
\newcommand{\Heiiuv}{He~{\sc ii} $\lambda$1640}
\newcommand{\Heii}{He~{\sc ii} $\lambda$4686}

\newcommand{\ebv}{\ifmmode E_{\rm B-V} \else $E_{\rm B-V}$\fi}
\newcommand{\av}{\ifmmode A_{\rm V} \else $A_{\rm V}$\fi}

\def\ergs{erg s$^{-1}$}

\def\msun{\ifmmode M_{\odot} \else M$_{\odot}$\fi}
\def\msunyr{\ifmmode M_{\odot} {\rm yr}^{-1} \else M$_{\odot}$ yr$^{-1}$\fi}
\def\zsun{\ifmmode Z_{\odot} \else Z$_{\odot}$\fi}

\def\lsun{\ifmmode L_{\odot} \else L$_{\odot}$\fi}

\def\mup{\ifmmode M_{\rm up} \else M$_{\rm up}$\fi}
\def\mlow{\ifmmode M_{\rm low} \else M$_{\rm low}$\fi}

\def\aap{A\&A}

\def\apjl{ApJL}
\def\apjs{ApJS}

\newcommand{\oh}{\ifmmode 12 + \log({\rm O/H}) \else$12 + \log({\rm O/H})$\fi}
\def\flyf{\ifmmode f_{\rm Lyf} \else $f_{\rm Lyf}$\fi}
\def\pz{\ifmmode P(z) \else $P(z)$\fi}
\def\ki2{\ifmmode \chi^2 \else $\chi^2$\fi}
\def\zphot{\ifmmode z_{\rm phot} \else $z_{\rm phot}$\fi}

\newcommand{\xphot}{\ifmmode x_\gamma \else $v_\gamma$\fi}
\newcommand{\xobs}{\ifmmode x_{\rm obs} \else $x_{\rm obs}$\fi}
\newcommand{\xcmf}{\ifmmode x_{\rm CMF} \else $x_{\rm CMF}$\fi}
\newcommand{\vexp}{\ifmmode V_{\rm exp} \else $V_{\rm exp}$\fi}
\newcommand{\vmax}{\ifmmode V_{\rm max} \else $V_{\rm max}$\fi}
\newcommand{\nh}{\ifmmode N_{\rm HI} \else $N_{\rm HI}$\fi}
\newcommand{\dv}{\ifmmode \Delta v({\rm em-abs}) \else $\Delta v({\rm em}-{\rm abs})$\fi}

\def\fesc{\ifmmode f_{\rm esc} \else $f_{\rm esc}$\fi}

\def\frellya{\ifmmode f^{\rm rel}_{\rm{Ly}\alpha} \else $f^{\rm rel}_{\rm{Ly}\alpha}$\fi}

\def\hii{H{\sc ii}}

\newcommand{\mstar}{\ifmmode M_\star \else $M_\star$\fi}
\newcommand{\muv}{\ifmmode M_{1500} \else $M_{1500}$\fi}
\newcommand{\auv}{\ifmmode A_{\rm UV} \else $A_{\rm UV}$\fi}
\newcommand{\luv}{\ifmmode L_{\rm UV} \else $L_{\rm UV}$\fi}
\newcommand{\lir}{\ifmmode L_{\rm IR} \else $L_{\rm IR}$\fi}
\newcommand{\lbol}{\ifmmode L_{\rm bol} \else $L_{\rm bol}$\fi}
\newcommand{\liruv}{\ifmmode L_{\rm IR+UV} \else $L_{\rm IR+UV}$\fi}
\newcommand{\liroveruv}{\ifmmode L_{\rm IR}/L_{\rm UV} \else $L_{\rm IR}/L_{\rm UV}$\fi}
\newcommand{\nlyc}{\ifmmode N_{\rm Lyc} \else $N_{\rm Lyc} $\fi}
\newcommand{\rholyc}{\ifmmode \rho_{\rm Lyc} \else $\rho_{\rm Lyc} $\fi}
\newcommand{\chion}{\ifmmode \xi_{\rm ion} \else $\xi_{\rm ion}$\fi}
\newcommand{\chioncorr}{\ifmmode \xi_{\rm ion}^0 \else $\xi_{\rm ion}^0$\fi}

\newcommand{\source}{SBS0335-052E}

\begin{document}

\title{X-ray binaries as the origin of nebular \heii\ emission in low-metallicity star-forming galaxies}
\subtitle{}
\author{D. Schaerer\inst{1,2}, 
T. Fragos\inst{1},
Y. I. Izotov$^{3}$,
}
  \institute{Observatoire de Gen\`eve, Universit\'e de Gen\`eve, 51 Ch. des Maillettes, 1290 Versoix, Switzerland
         \and
CNRS, IRAP, 14 Avenue E. Belin, 31400 Toulouse, France
        \and
Bogolyubov Institute for Theoretical Physics,
National Academy of Sciences of Ukraine, 14-b Metrolohichna str., Kyiv,
03143, Ukraine
         }

\authorrunning{D.\ Schaerer et al.}
\titlerunning{Nebular \heii\ emission}

\date{Received 2 January 2019 / Accepted 17 January 2019}


\abstract{The origin of nebular \heii\ emission, which is frequently observed in low-metallicity (O/H) star-forming galaxies,
remains largely an unsolved question. 
Using the observed anticorrelation of the integrated X-ray luminosity per unit of star formation rate ($L_X/{\rm SFR}$) of an X-ray binary 
population with metallicity and other empirical data from the well-studied galaxy I Zw 18, 
we show that the observed \Heii\ intensity and its trend with metallicity is naturally reproduced if the bulk of He$^+$ ionizing 
photons are emitted by the X-ray sources.
We also show that  a combination of X-ray binary population models with normal single and/or binary stellar models
reproduces the observed $I(4686)/I(\hb)$ intensities and its dependency on metallicity and age.
We conclude that both empirical data and theoretical models suggest that high-mass X-ray binaries are 
the main source of nebular \heii\ emission in low-metallicity star-forming galaxies.
}


 \maketitle

\section{Introduction}
\label{s_intro}
Since its discovery, the origin of nebular \Heii\ emission observed in (giant) \hii\ regions, \hii\ galaxies, dwarf galaxies, Wolf-Rayet galaxies,
and similar sources that are known or generally thought to be powered by stellar radiation 
\citep[e.g.,][]{Pakull1986Detection-of-an,Garnett1991He-II-emission-,Guseva2000A-Spectroscopic,Shirazi2012Strongly-star-f}
has remained mysterious.
The nebular \heii\ emission lines (the \Heii\ or the \Heiiuv\ UV line) 
in the integrated spectrum indicate the existence of sources
of hard ionizing radiation, that is,\ photons with energies above 54 eV (or $\lambda < 228$ \AA, corresponding to the ionization potential
of \hep).

Only stars with very high effective temperatures $\teff \ga 80-100$ kK emit non-negligible amounts of \hep\
ionizing photons \citep[e.g.,][]{Tumlinson2000Zero-Metallicit,schaerer2002}.  Because such temperatures are only reached in very peculiar evolutionary 
phases (e.g.,\ in the Wolf-Rayet (WR) or planetary nebula phase), ``normal'' stellar populations contain few such stars. Their ionizing spectra are therefore in general not hard enough to explain the observed nebular \Heii\ emission, with typical intensities $I(4686)/I(\hb)$ 
of a few percent  \citep[e.g.,][]{Izotov2004Systematic-Effe}.
Exceptions are populations of very low metallicity ($\la 10^{-4}$ solar) or zero metallicity \citep[PopIII; cf.][]{schaerer2003}, 
which are not observed, however. 
Alternatively, the observed nebular \heii\ emission can be explained by the presence of WR stars in some galaxies
with metallicities $\oh \ga 8.4$ \citep[e.g.,][]{Schaerer1996About-the-Initi,Shirazi2012Strongly-star-f}.
However, only a small fraction of nebular \heii\ emitters show WR stars, and the intensity of \Heii\ increases with decreasing
metallicity (cf.\ also below), opposite to that of the WR populations \citep{Guseva2000A-Spectroscopic}. 
For all these reasons, photoionization models using ionizing spectra predicted from  ``standard'' stellar populations
fail to reproduce the observed nebular \heii\ emission, especially at low metallicities, $\oh \la 8.2$, where it is most prominent 
\citep[cf.][]{Shirazi2012Strongly-star-f,Stasinska2015Excitation-prop}.

Several studies have explored the impact of various physical processes, such as stellar rotation and binarity, 
on the evolution of massive stars, predicting for instance\ the existence of hotter single main-sequence stars in cases of very fast 
rotation leading to quasi-homogeneous evolution \citep[see, e.g.,][]{Szecsi2015Low-metallicity,Maeder1987Evidences-for-a},
and ``rejuvenated'', hot stars formed in interacting binary systems \citep[e.g.,][]{van-Bever1998The-rejuvenatio,Eldridge2017Binary-Populati,Gotberg2018Spectral-models},
both of which  imply hotter stellar populations on average and hence a harder ionizing spectrum.
However, none of these works has so far been able to quantitatively explain the observed intensity of the 
\heii\ emission in low-metallicity galaxies. For example, the latest BPASS binary population and spectral synthesis models 
incorporating the effect of binary mass transfer on stellar evolution falls short of predicting the observed \Heii/\hb\ intensities
by typically one order of magnitude, even considering different stellar initial mass functions
\citep[see][and below]{Stanway2018Initial-Mass-Fu}.

Since the first discoveries of \heii\ emitters, other sources and processes that could emit more energetic photons 
than stellar sources have been suggested. These include\ X-ray binaries (XRBs),  photoionization by X-rays, and strong shocks.
X-rays  can explain specific cases, but appear insufficient in others
\citep[e.g.,][]{Pakull1986Detection-of-an,Garnett1991He-II-emission-,Thuan2005High-Ionization,Kehrig2015The-Extended-He}.
For some galaxies it is argued that shocks can explain \heii\ emission, if they provide $\sim 10$ \% of the hydrogen ionizing photons
\citep[e.g.,][]{Thuan2005High-Ionization,Izotov2012The-detection-o}.
However, no straightforward predictive model that would allow linking shock models to other galaxy properties exists, 
and it is not clear how shocks would reproduce the observed trends of \heii\ intensity with metallicity (cf.\ below).
In short, the origin of nebular \heii\ emission in star-forming galaxies is so far considered an unsolved question,
with numerous important implications, also for our understanding of distant galaxies
\citep[e.g.,][]{Cassata2013He-II-emitters-,Stark2016Galaxies-in-the,Steidel2016Reconciling-the,Vanzella2016High-resolution,Berg2018A-Window-on-the}.

Here we present a new approach to this question, considering observational and empirical findings from X-rays and optical spectroscopy,
insight from detailed modeling of the low-metallicity galaxy I Zw 18, and recent models of XRB populations.
We examine in particular the implications of the increased importance of X-ray emission in low-metallicity galaxies
\citep[e.g.,][]{Kaaret2011X-Rays-from-Blu,Basu-Zych2013Evidence-for-El,Douna2015Metallicity-dep,Brorby2016Enhanced-X-ray-}
on the \heii\ problem.
We show that massive X-binaries and/or ultra-luminous X-rays sources (ULX), which are a natural extension of the latter to 
higher X-ray luminosities, are able to reproduce the basic observed trends and are therefore the most likely sources of 
nebular \Heii\ emission in low-metallicity star-forming galaxies.

\section{Comparing observed scalings of nebular \Heii\ emission and X-rays }
\label{s_obs}

\subsection{Star-forming galaxy samples}
Low-metallicity  ($\oh \la 8.4$) galaxies, where nebular \heii\ is frequently detected, exhibit one main observational
behavior that we show in Fig.\  1: 
a clear trend of increasing line intensity $I(4686)/I(\hb)$ with decreasing metallicity.
To illustrate the \heii\ observations, we here use a compilation of more than 1400 star-forming galaxies or regions thereof
from \cite{Izotov2016The-bursting-na}, which are selected for the quality of the spectra, allowing direct metallicity determinations
via the classical auroral line method\footnote{The
sources include 1135 sources from the Sloan Digital Sky Survey (SDSS), and 328 are from observations obtained by Izotov
and collaborators with 4-8m class telescopes. Standard emission line ratios were used to select star-forming galaxies, not AGN.}.
The $I(4686)/I(\hb)$ detections follow a sub-linear scaling with metallicity (with a slope of $\sim -0.45$), as indicated by the yellow line.
Our aim is not to properly determine a mean relation for this quantity taking completeness and other factors\ into account.
Our goal is instead to show how ``typical'' sources and observed trends with metallicity can be reproduced.

Interestingly, the analysis of the X-ray emission of star-forming galaxies shows that the X-ray luminosity, $L_X$, 
normalized per  star formation rate (SFR), also increases with decreasing metallicity with a similar dependence 
as the \heii\ intensity. \cite{Douna2015Metallicity-dep} obtained a near-linear scaling of $L_X/{\rm SFR}$ with O/H,
whereas \cite{Brorby2016Enhanced-X-ray-} found a somewhat shallower dependence. Both studies showed an increase of 
$L_X/{\rm SFR}$ by more than one order of magnitude from solar metallicity to the lowest metallicity X-ray detected galaxies, 
such as I Zw 18 at $\sim 1/30$ solar \citep[\oh = 7.18,][]{izotov98}.

The empirical relations between $L_X/{\rm SFR}$  and O/H can easily be translated into a
relation between the  $I(4686)/I(\hb)$ intensity ratio and O/H.  We only need to assume or determine one parameter,
the average He$^+$ ionizing photon flux per $L_X$ of the sources responsible for the X-ray luminosity 
\begin{equation}
        q = Q({\rm He}^+)/L_X.
\label{eq_q2}
\end{equation}
Given the proportionality between the Lyman continuum photon flux and the SFR, 
that is,\ taking $Q({\rm H})/{\rm SFR}=9.26 \times 10^{52}$ photon s$^{-1}$/(\msunyr) \citep{kennicutt1998}, 
this allows us to derive the ratio of He$^+$ over H ionizing photons, $Q({\rm He}^+)/Q({\rm H})$, 
which directly yields the observed \Heii\ line intensity
\begin{equation}
        I(4686)/I(\hb) = A \times Q({\rm He}^+)/Q({\rm H}),
\label{eq_heii}
\end{equation}
where $A \approx 1.74$ for typical nebular conditions \citep[see, e.g.,][]{Stasinska2015Excitation-prop}.
Here we have assumed that the SFR is constant over $\ga 10$ Myr, which is consistent with the 
assumption of even longer SF timescales ($\ga 100$ Myr) made for the empirical SFR determinations
that were used by the above studies in determining $L_X/{\rm SFR}$.

The \heii\ intensity as a function of metallicity predicted from the empirical $L_X/{\rm SFR}$ versus O/H relations
of \cite{Douna2015Metallicity-dep} and \cite{Brorby2016Enhanced-X-ray-} assuming $q=2\times 10^{10}$ photon/erg
(cf.\ below) are shown in Fig.\ 1.
We suggest that the close similarity of the dependence of $L_X/{\rm SFR}$ and $I(4686)/I(\hb)$ on metallicity
also indicates a causal connection between them, that is to say, that X-ray sources are the dominant source
of He$^+$ ionizing photons.
We now examine I Zw18, one of the best-studied low-metallicity galaxies, in particular, to estimate 
$q$ empirically.

\setcounter{figure}{0}
\begin{figure}[tb]
{\centering
\includegraphics[width=9cm]{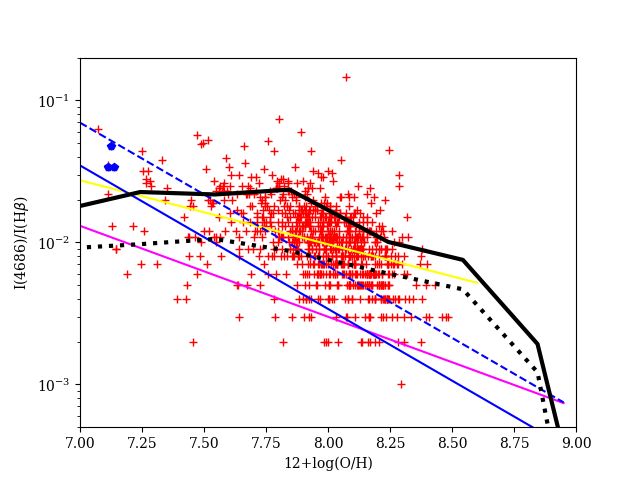}
\caption{Observed and predicted $I(4686)/I(\hb)$ relative nebular line intensities as a function of metallicity.
Observations of low-metallicity star-forming galaxies are shown as  red plusses, 
and different measurements of the NW region of I Zw 18 with blue diamonds.
The yellow line shows the linear regression to the data points.
Assuming a default value of $q=2\times 10^{10}$ photon/erg, the empirical $L_X/{\rm SFR}$--O/H relations of 
\cite{Douna2015Metallicity-dep} and  \cite{Brorby2016Enhanced-X-ray-}  translate into \heii\ intensities 
shown by the blue and magenta solid lines, respectively.
The black lines show the predicted \heii\ intensity adopting $L_X/{\rm SFR}$ predicted from 
the XRB synthesis models described in Sect.\ \protect\ref{s_models} 
for a constant SFR over 10 Myr (dotted) and 0.1 Gyr (solid), and the same value of $q$.
The blue dashed line differs from the solid line by assuming a value of $q$ that is a factor of two higher.
}}
\label{fig_heii_oh}
\end{figure}

\subsection{I Zw 18 as a testbed}
The low-metallicity dwarf galaxy I Zw 18 has been studied
in detail. It is known to show nebular \heii\ emission concentrated in the NW region with an integrated intensity of  
$I(4686)/I(\hb) \sim 0.02-0.04$ \citep[e.g.,][]{izotov98,Kehrig2015The-Extended-He,Lebouteiller2017Neutral-gas-hea}.
The NW region, a giant \hii\ region, also harbors a point-like X-ray source that dominates the observed X-ray emission 
of this galaxy and is suggested to be a single XRB \citep{Thuan2004Chandra-Observa,Kaaret2013A-State-Transit}.
A point-source luminosity $L_X=3. \times 10^{39}$ \ergs\ in the 0.5-10 keV range has been determined by 
\cite{Thuan2004Chandra-Observa} from Chandra observations, whereas \cite{Kaaret2013A-State-Transit} find
a higher $L_X=1.4 \times 10^{40}$ \ergs\ from XMM data, possibly indicating variability according to these authors.

Using the vast set of multiwavelength observations, \cite{Lebouteiller2017Neutral-gas-hea} have recently constructed
a tailored model of the neutral and ionized interstellar matter (ISM) of this galaxy.
Based on their detailed photoionization modeling including stellar radiation, emission from an accretion disk model
that successfully fit the X-ray source, and cosmic rays, they were able to reproduce all the observational constraints.
Their model shows in particular that both the emission lines of low- and high-ionization stages, including also 
\heii, which are strongly affected by high-energy radiation, are well explained by the observed X-ray source in this galaxy.
A successful physical model including ionization from a stellar black hole was also constructed by \cite{Heap2019Radiative-signa}.
From the observed properties of I Zw 18, we can therefore empirically derive the amount of He$^+$ ionizing photons 
emitted from the X-ray source, assuming that the latter dominates the ionizing photon flux above 54 eV.

Using the observations of  value of \cite{Kehrig2015The-Extended-He}, who find $Q({\rm He}^+)=1.33 \times 10^{50}$ photon s$^{-1}$,
and the two values of the X-ray luminosities reported above, we thus obtain
$q= Q({\rm He}^+)/L_X=(1.0-3.4)\times 10^{10}$ photon/erg. Adopting an intermediate value of $q=2\times 10^{10}$ photon/erg,
we find that the observed $I(4686)/I(\hb)$ intensity of I Zw 18 is well reproduced from the $L_X/{\rm SFR}$ relation of 
\cite{Douna2015Metallicity-dep}, as shown in Fig.\ 1.
This is indeed expected because I Zw 18 lies perfectly
on their mean relation.  However, the bulk of the observations are better reproduced with a higher value of 
$q=4\times 10^{10}$ photon/erg, as also shown in this figure.

\section{Nebular \Heii\ emission from X-ray binary population models}
\label{s_models}

We now explore how XRB population synthesis models compare with the nebular \Heii\ observations.
To do this we use models developed by \citet{Fragos2013X-Ray-Binary-Ev} and \cite{Fragos2013Energy-Feedback}
to study the cosmological evolution of XRB populations that were recently recalibrated to updated measurements of the cosmic star-formation history and metallicity evolution \citep{Madau2017Radiation-Backg}.
The same type of models and choice of model parameters has been shown to produce XRB populations that are consistent with observation of both the local and the distant universe 
\citep[e.g.,][]{Tzanavaris2013Modeling-X-Ray-,Tremmel2013Modeling-the-Re,Lehmer2016The-Evolution-o}.
%
An important feature of the models is the strong dependence of the XRB population on metallicity, both
 in terms of the formation efficiency of XRBs and the integrated  X-ray luminosity of the whole population.

The predictions of the integrated X-ray luminosity per unit of SFR from the synthetic models as a function of stellar population age and metallicity for simple stellar populations (bursts) are shown in Fig. 2. The X-ray luminosity corresponds to the $0.3-10$ keV range and absorption from a hydrogen column density of $N_{\rm H}=3\times 10^{21}$ cm$^{-2}$  \citep{Madau2017Radiation-Backg}.

Assuming, for example, SFR=const over a period of 0.1 Gyr, we find an increase of $I(4686)/I(\hb)$ by more
than order of magnitude between solar metallicity and $\oh \sim 7.6$, and a flattening thereafter, as shown in Fig.\ 1.
For shorter periods of constant SFR the result is very similar; for longer periods the \heii\ intensity may be
somewhat higher. In any case, the models follow both the observed  line intensity and its metallicity dependence quite well.
Here the predicted values of $L_X/{\rm SFR}$ have been translated into the observed $I(4686)/I(\hb)$ intensity using the 
same simple assumptions we described above, that is,\ with one adjusted parameter ($q$).

Because the \heii\ intensity is also expected to depend on the age of stellar population, we examine in Fig.\ 3
its dependence on the \hb\ equivalent width, $W(\hb)$, a well-known proxy for age in young simple stellar populations (SSPs).
To do so we combine the predictions from the BPASSv2.1 synthesis models, describing the evolution of an ensemble of 
single and binary stars as a function of age, with that of the XRB population resulting
from a stellar population with the same total mass. BPASS yields the predicted temporal evolution of $Q(H)$ and $W(\hb)$, 
whereas the XRB model predicts $L_X(t)$, from which derive $Q({\rm He}^+)$ from Eq.\ \ref{eq_q2}, hence $I(4686)/I(\hb)$.

As shown in Fig.\ 3,
the SSP models including X-ray binaries show an increase of $I(4686)/I(\hb)$ by
more than one order of magnitude compared to the BPASS models, which include interacting binaries, but not the XRB phase.
The models cover the range of observed \heii\ intensities and are in fair agreement with most of the data. 
Possible disagreements remain for the galaxies with the highest $W(\hb)\ga 200$ \AA, which correspond to ages younger 
than $\la 5$ Myr according to the BPASS models used here. This would suggest that luminous X-ray sources  appear 
fairly soon to explain \heii\ emission from these galaxies as well.
This is consistent with detailed binary evolutionary models, which show that the first ULX sources may turn 
on as early as $\sim 4$ Myr after a burst of star formation \citep[e.g., see Fig.\ 4 in ][]{Rappaport2005Stellar-mass-bl}.
Alternatively, X-ray sources from an underlying older population (e.g.,\ with a constant SFR,
cf.\ Fig.\ 1), or shocks  \citep[cf.][]{Izotov2012The-detection-o,Izotov2019J12343901:-an-e}
could  also contribute to boost the \heii\ emission in these sources.

We conclude that when we include the contribution of XRBs in our stellar population models and making a simple assumption on
the He$^+$ ionizing photon flux emitted by the X-ray sources, we are able to generally reproduce the observed 
nebular \heii\ emission in star-forming galaxies at low metallicity,  although some difficulties may exist 
in explaining the youngest sources.

\begin{figure}[tb]
{\centering
\includegraphics[width=9cm]{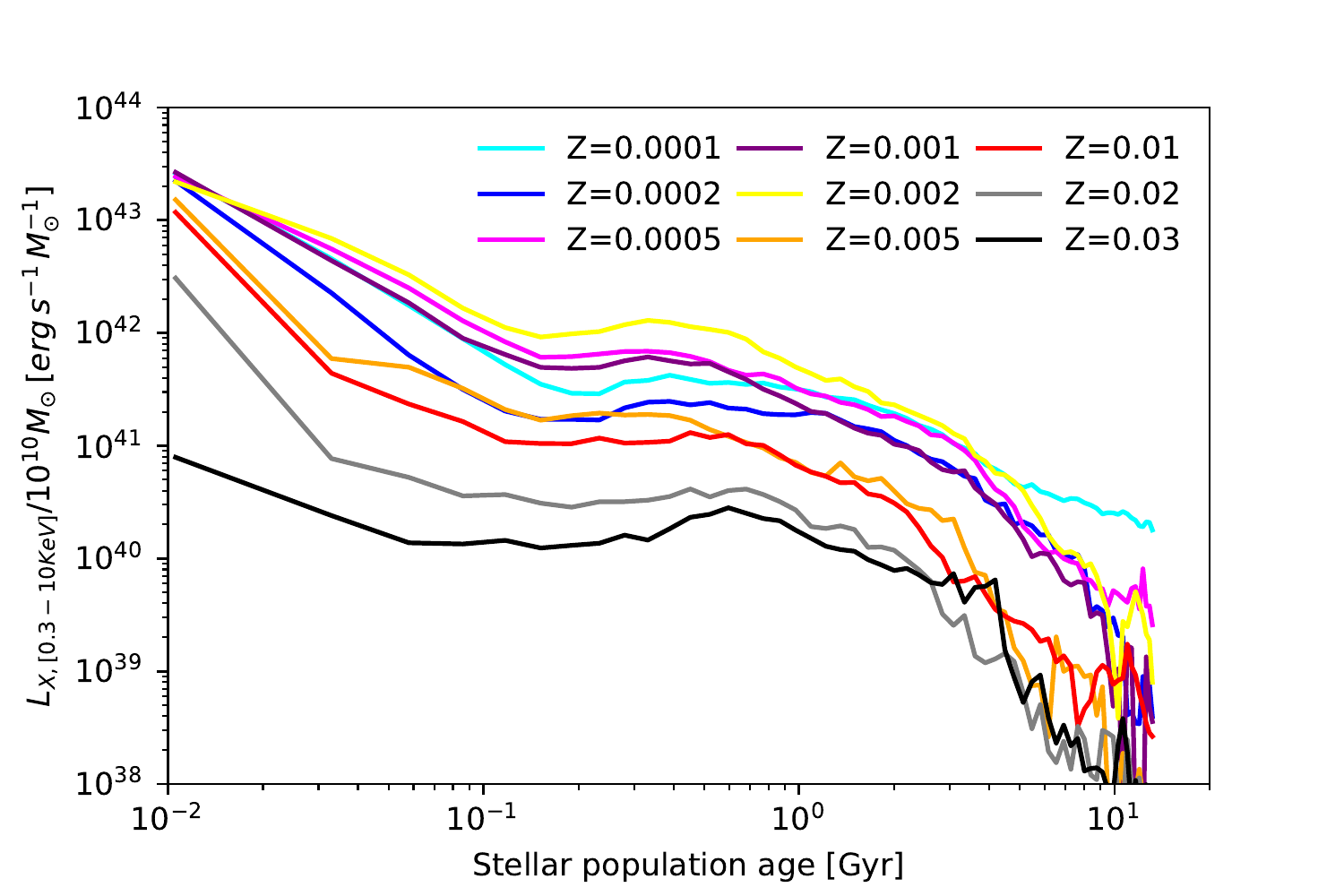}
\caption{Predicted X-ray luminosity as a function of age for simple stellar populations with different metallicities.}
}
\label{fig_lx_age}
\end{figure}

\begin{figure}[tb]
{\centering
\includegraphics[width=9cm]{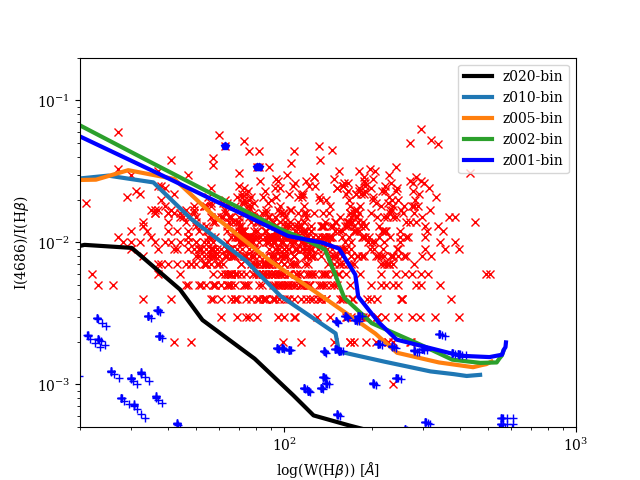}
\caption{Observed and predicted $I(4686)/I(\hb)$ relative nebular line intensities as a function of the \hb\ equivalent width.
Observations are the same as in Fig.\ 1.
Blue crosses show the predictions from the BPASS models of \protect\cite{Xiao2018Emission-line-d}, 
which fail to reproduce the observed \Heii\ intensities. 
Predictions for simple stellar populations of different metallicities (shown by the solid lines), derived from the combination of BPASS + XRB models,
show a fair agreement with the observations.
}
}
\label{fig_heii_whb}
\end{figure}

\section{Discussion}
\label{s_discuss}

\subsection{Non-detections and scatter}
In our compilation, approximately two-thirds of the spectra show nebular \Heii\ emission, with no indication for a
dependence of the detection fraction on metallicity.
While the empirical scaling relations and the simple models described above
explain the presence and strength of \heii, we also need to consider the cases of \Heii\ of non-detections.
There are many possible explanations for this, for example, stochasticity effects related to 
the XRB population, variability of the X-ray sources,  and variations in the assumed mean value of $q$ (Eq.\ \ref{eq_q2}).
Stochasticity related to the low SFR of the observed low-metallicity galaxies also explains the observed scattering 
in X-rays, as shown by \cite{Douna2015Metallicity-dep}.
Alternatively, for a given ratio $Q({\rm He}^+)/Q({\rm H})$, the  $I(4686)/I(\hb)$ intensity may be lower
than predicted here (assuming Case B) in nebulae with a low-ionization parameter,
as discussed by \cite{raiter2010} and first explained by \cite{Stasinska1986Intermediate-ma}.

Some cases of strong \heii\ emission but faint X-ray emission are also known.
Such a prominent case is the well-studied low-metallicity galaxy \source, for example, which shows extended \Heii\ emission
with an integrated intensity $I(4686)/I(\hb) \approx 0.026$, but faint X-ray emission  
\citep[see][]{Kehrig2018The-extended-He,Thuan2004Chandra-Observa}.
The observed value of $q \sim (5-10) \times 10^{11}$ photon/erg for \source\ is thus significantly higher 
($\sim 25-50$ times) than the mean value adopted here, and higher than what is observed in I Zw 18. This casts some doubt on the responsibility of X-rays for the observed \heii\ emission \citep{Kehrig2018The-extended-He}.
Interestingly, however, the bulk of the X-ray emission observed with Chandra in \source\ is due to a point-like source that
is spatially coincident with the brightest super star clusters and the main emission region of \Heii.
We suggest that beamed X-ray emission could explain the relatively low observed X-ray flux, but produce the
observed \heii\ emission if the He$^+$ ionizing photons are absorbed relatively close to the source.
If correct, this would imply an intrinsic $L_X \sim (0.8-1.6) \times 10^{41}$ \ergs, corresponding to several 
high-mass XRBs or a single ULX.
ULXs, and especially those with neutron star accretors, which are expected to slightly dominate older populations 
(mean stellar population age $> 30$ Myr), are expected to be highly beamed source
\citep[e.g.,][]{King2009Masses-beaming-,Wiktorowicz2018The-observed-vs},
and thus the observed X-ray luminosity per unit of star formation can be a significant underestimate in certain cases. 

\subsection{Improvements and future tests}
Although our results show that with the assumption of one parameter $q$, the amount of He$^+$ ionizing photons 
emitted per X-ray luminosity, the high-mass XRB / ULX population can reproduce both the observed 
nebular \heii\ emission in star-forming galaxies and its metallicity dependence, it is difficult
to definitely prove  the validity of our explanation.
Further detailed multiwavelength studies of individual objects, as done for I Zw 18 by \cite{Lebouteiller2017Neutral-gas-hea},
could provide new insight. 
Clearly, a next step is to examine in depth the correlation between \heii\ and X-ray emission on
an individual object basis, and for the largest possible samples.
Future studies should ideally also attempt to predict the spectrum of the XRBs and their
emission at low energies to allow more quantitative work that also examines their possible contribution 
to other emission lines, which are due to stellar photoionization.

Although we suggest that XRBs are the main source of nebular \Heii\ emission at low metallicity, we do not
exclude contributions from shocks, as advocated for some galaxies \citep[e.g.,][]{Thuan2005High-Ionization,Izotov2012The-detection-o}.
Establishing their role for larger samples and distinguishing between shock and XRB contributions should help to
refine our understanding of the high-energy sources in metal-poor star-forming galaxies.

\section{Conclusion}
\label{s_conclude}
We have combined results from the detailed multiwavelength study of I Zw 18 from \cite{Lebouteiller2017Neutral-gas-hea},
which shows that nebular \Heii\ emission in this low-metallicity galaxy can be explained by an observed massive XRB,
with the observed anticorrelation of $L_X/{\rm SFR}$ with metallicity \citep{Douna2015Metallicity-dep,Brorby2016Enhanced-X-ray-}.
With this we showed that nebular \heii, which is frequently observed in low-metallicity star-forming galaxies, is most likely due to
high-mass XRBs and/or ULX.

Assuming that the bulk of the hydrogen ionizing photons is emitted by normal stars (single or binaries) and that
the photons above 54 eV (capable of ionizing He$^+$) are emitted by X-ray binaries, we have shown that the 
observed anticorrelation of $L_X/{\rm SFR}$ with O/H reproduces the observed dependence of $I(4686)/I(\hb)$
on metallicity well (Fig.\ 1). A single parameter $q$, which describes the ratio of He$^+$ ionizing photons per X-ray luminosity
emitted from the XRB population and is determined from I Zw 18, suffices to describe this behavior.

To further validate our estimates, we have used the recently recalibrated \citep{Madau2017Radiation-Backg} XRB population synthesis models 
of \citet{Fragos2013X-Ray-Binary-Ev,Fragos2013Energy-Feedback},
which predict a strong metallicity dependence  of the X-ray emission to examine their implications for nebular \heii\ emission.
With the same assumptions, we found that these models also predict both the observed $I(4686)/I(\hb)$ intensity
and its dependence on O/H for constant SFRs (Fig.\ 1).
Finally, we predicted the age-dependent \heii\ emission of simple stellar populations at different metallicities
by combining the BPASS binary models \citep{Eldridge2017Binary-Populati,Xiao2018Emission-line-d} with the XRB models.
Our predictions are in good agreement with the bulk of the observations (see Fig.\ 3). 
In some galaxies with very high \hb\ equivalent widths ($W(\hb)\ga 200$ \AA), the observed \heii\ emission is stronger
than predicted, which may indicate that luminous X-ray sources should appear on very short timescales ($\la 5$ Myr).

Overall, we conclude that both empirical data and theoretical models suggest that
high-mass X-ray binaries are the main source of nebular \heii\ emission in low-metallicity star-forming galaxies,
although contributions from shocks are not excluded.

\begin{acknowledgements}
DS wishes  to thank Vianney Lebouteiller for exchanges on his modeling of I Zw 18.
TF is grateful for support from the SNSF Professorship grant (project number PP00P2176868) 
Y.I.\ acknowledges support from the National Academy of Sciences of Ukraine (Project No.\ 0116U003191) and by its
Program of Fundamental Research of the Department of Physics and Astronomy (Project No.\ 0117U000240).

\end{acknowledgements}
\bibliographystyle{aa}
\bibliography{merge_misc_highz_literature}

\begin{thebibliography}{49}
\expandafter\ifx\csname natexlab\endcsname\relax\def\natexlab#1{#1}\fi

\bibitem[{{Basu-Zych} {et~al.}(2013){Basu-Zych}, {Lehmer}, {Hornschemeier},
  {Gon{\c c}alves}, {Fragos}, {Heckman}, {Overzier}, {Ptak}, \&
  {Schiminovich}}]{Basu-Zych2013Evidence-for-El}
{Basu-Zych}, A.~R., {Lehmer}, B.~D., {Hornschemeier}, A.~E., {et~al.} 2013,
  \apj, 774, 152

\bibitem[{{Berg} {et~al.}(2018){Berg}, {Erb}, {Auger}, {Pettini}, \&
  {Brammer}}]{Berg2018A-Window-on-the}
{Berg}, D.~A., {Erb}, D.~K., {Auger}, M.~W., {Pettini}, M., \& {Brammer}, G.~B.
  2018, \apj, 859, 164

\bibitem[{{Brorby} {et~al.}(2016){Brorby}, {Kaaret}, {Prestwich}, \&
  {Mirabel}}]{Brorby2016Enhanced-X-ray-}
{Brorby}, M., {Kaaret}, P., {Prestwich}, A., \& {Mirabel}, I.~F. 2016, \mnras,
  457, 4081

\bibitem[{{Cassata} {et~al.}(2013){Cassata}, {Le F{\`e}vre}, {Charlot},
  {Contini}, {Cucciati}, {Garilli}, {Zamorani}, {Adami}, {Bardelli}, {Le Brun},
  {Lemaux}, {Maccagni}, {Pollo}, {Pozzetti}, {Tresse}, {Vergani}, {Zanichelli},
  \& {Zucca}}]{Cassata2013He-II-emitters-}
{Cassata}, P., {Le F{\`e}vre}, O., {Charlot}, S., {et~al.} 2013, \aap, 556, A68

\bibitem[{{Douna} {et~al.}(2015){Douna}, {Pellizza}, {Mirabel}, \&
  {Pedrosa}}]{Douna2015Metallicity-dep}
{Douna}, V.~M., {Pellizza}, L.~J., {Mirabel}, I.~F., \& {Pedrosa}, S.~E. 2015,
  \aap, 579, A44

\bibitem[{{Eldridge} {et~al.}(2017){Eldridge}, {Stanway}, {Xiao}, {McClelland},
  {Taylor}, {Ng}, {Greis}, \& {Bray}}]{Eldridge2017Binary-Populati}
{Eldridge}, J.~J., {Stanway}, E.~R., {Xiao}, L., {et~al.} 2017, \pasa, 34, e058

\bibitem[{{Fragos} {et~al.}(2013{\natexlab{a}}){Fragos}, {Lehmer}, {Tremmel},
  {Tzanavaris}, {Basu-Zych}, {Belczynski}, {Hornschemeier}, {Jenkins},
  {Kalogera}, {Ptak}, \& {Zezas}}]{Fragos2013X-Ray-Binary-Ev}
{Fragos}, T., {Lehmer}, B., {Tremmel}, M., {et~al.} 2013{\natexlab{a}}, \apj,
  764, 41

\bibitem[{{Fragos} {et~al.}(2013{\natexlab{b}}){Fragos}, {Lehmer}, {Naoz},
  {Zezas}, \& {Basu-Zych}}]{Fragos2013Energy-Feedback}
{Fragos}, T., {Lehmer}, B.~D., {Naoz}, S., {Zezas}, A., \& {Basu-Zych}, A.
  2013{\natexlab{b}}, \apjl, 776, L31

\bibitem[{{Garnett} {et~al.}(1991){Garnett}, {Kennicutt}, {Chu}, \&
  {Skillman}}]{Garnett1991He-II-emission-}
{Garnett}, D.~R., {Kennicutt}, Jr., R.~C., {Chu}, Y.-H., \& {Skillman}, E.~D.
  1991, \apj, 373, 458

\bibitem[{{G{\"o}tberg} {et~al.}(2018){G{\"o}tberg}, {de Mink}, {Groh},
  {Kupfer}, {Crowther}, {Zapartas}, \& {Renzo}}]{Gotberg2018Spectral-models}
{G{\"o}tberg}, Y., {de Mink}, S.~E., {Groh}, J.~H., {et~al.} 2018, \aap, 615,
  A78

\bibitem[{{Guseva} {et~al.}(2000){Guseva}, {Izotov}, \&
  {Thuan}}]{Guseva2000A-Spectroscopic}
{Guseva}, N.~G., {Izotov}, Y.~I., \& {Thuan}, T.~X. 2000, \apj, 531, 776

\bibitem[{{Heap} {et~al.}(2019){Heap}, {Hubeny}, {Bouret}, \&
  {Lanz}}]{Heap2019Radiative-signa}
{Heap}, S.~R., {Hubeny}, I., {Bouret}, J.-C., \& {Lanz}, T. 2019, in Radiative
  signatures from the cosmos, ed. K.~{Werner} \& T.~{Rauch}, ASP Conference
  Series, in press

\bibitem[{{Izotov} {et~al.}(2016){Izotov}, {Guseva}, {Fricke}, \&
  {Henkel}}]{Izotov2016The-bursting-na}
{Izotov}, Y.~I., {Guseva}, N.~G., {Fricke}, K.~J., \& {Henkel}, C. 2016,
  \mnras, 462, 4427

\bibitem[{{Izotov} \& {Thuan}(1998)}]{izotov98}
{Izotov}, Y.~I. \& {Thuan}, T.~X. 1998, \apj, 500, 188

\bibitem[{{Izotov} \& {Thuan}(2004)}]{Izotov2004Systematic-Effe}
{Izotov}, Y.~I. \& {Thuan}, T.~X. 2004, \apj, 602, 200

\bibitem[{{Izotov} {et~al.}(2019){Izotov}, {Thuan}, \&
  {Guseva}}]{Izotov2019J12343901:-an-e}
{Izotov}, Y.~I., {Thuan}, T.~X., \& {Guseva}, N.~G. 2019, \mnras, 483, 5491

\bibitem[{{Izotov} {et~al.}(2012){Izotov}, {Thuan}, \&
  {Privon}}]{Izotov2012The-detection-o}
{Izotov}, Y.~I., {Thuan}, T.~X., \& {Privon}, G. 2012, \mnras, 427, 1229

\bibitem[{Kaaret \& Feng(2013)}]{Kaaret2013A-State-Transit}
Kaaret, P. \& Feng, H. 2013, The Astrophysical Journal, 770, 20

\bibitem[{{Kaaret} {et~al.}(2011){Kaaret}, {Schmitt}, \&
  {Gorski}}]{Kaaret2011X-Rays-from-Blu}
{Kaaret}, P., {Schmitt}, J., \& {Gorski}, M. 2011, \apj, 741, 10

\bibitem[{{Kehrig} {et~al.}(2018){Kehrig}, {V{\'{\i}}lchez}, {Guerrero},
  {Iglesias-P{\'a}ramo}, {Hunt}, {Duarte-Puertas}, \&
  {Ramos-Larios}}]{Kehrig2018The-extended-He}
{Kehrig}, C., {V{\'{\i}}lchez}, J.~M., {Guerrero}, M.~A., {et~al.} 2018,
  \mnras, 480, 1081

\bibitem[{{Kehrig} {et~al.}(2015){Kehrig}, {V{\'{\i}}lchez},
  {P{\'e}rez-Montero}, {Iglesias-P{\'a}ramo}, {Brinchmann}, {Kunth}, {Durret},
  \& {Bayo}}]{Kehrig2015The-Extended-He}
{Kehrig}, C., {V{\'{\i}}lchez}, J.~M., {P{\'e}rez-Montero}, E., {et~al.} 2015,
  \apjl, 801, L28

\bibitem[{{Kennicutt}(1998)}]{kennicutt1998}
{Kennicutt}, Jr., R.~C. 1998, \araa, 36, 189

\bibitem[{{King}(2009)}]{King2009Masses-beaming-}
{King}, A.~R. 2009, \mnras, 393, L41

\bibitem[{{Lebouteiller} {et~al.}(2017){Lebouteiller}, {P{\'e}quignot},
  {Cormier}, {Madden}, {Pakull}, {Kunth}, {Galliano}, {Chevance}, {Heap},
  {Lee}, \& {Polles}}]{Lebouteiller2017Neutral-gas-hea}
{Lebouteiller}, V., {P{\'e}quignot}, D., {Cormier}, D., {et~al.} 2017, \aap,
  602, A45

\bibitem[{{Lehmer} {et~al.}(2016){Lehmer}, {Basu-Zych}, {Mineo}, {Brandt},
  {Eufrasio}, {Fragos}, {Hornschemeier}, {Luo}, {Xue}, {Bauer}, {Gilfanov},
  {Ranalli}, {Schneider}, {Shemmer}, {Tozzi}, {Trump}, {Vignali}, {Wang},
  {Yukita}, \& {Zezas}}]{Lehmer2016The-Evolution-o}
{Lehmer}, B.~D., {Basu-Zych}, A.~R., {Mineo}, S., {et~al.} 2016, \apj, 825, 7

\bibitem[{{Madau} \& {Fragos}(2017)}]{Madau2017Radiation-Backg}
{Madau}, P. \& {Fragos}, T. 2017, \apj, 840, 39

\bibitem[{{Maeder}(1987)}]{Maeder1987Evidences-for-a}
{Maeder}, A. 1987, \aap, 178, 159

\bibitem[{Pakull \& Angebault(1986)}]{Pakull1986Detection-of-an}
Pakull, M.~W. \& Angebault, L.~P. 1986, Nature, 322, 511 EP

\bibitem[{{Raiter} {et~al.}(2010){Raiter}, {Schaerer}, \&
  {Fosbury}}]{raiter2010}
{Raiter}, A., {Schaerer}, D., \& {Fosbury}, R.~A.~E. 2010, \aap, 523, A64

\bibitem[{{Rappaport} {et~al.}(2005){Rappaport}, {Podsiadlowski}, \&
  {Pfahl}}]{Rappaport2005Stellar-mass-bl}
{Rappaport}, S.~A., {Podsiadlowski}, P., \& {Pfahl}, E. 2005, \mnras, 356, 401

\bibitem[{{Schaerer}(1996)}]{Schaerer1996About-the-Initi}
{Schaerer}, D. 1996, \apjl, 467, L17

\bibitem[{{Schaerer}(2002)}]{schaerer2002}
{Schaerer}, D. 2002, \aap, 382, 28

\bibitem[{{Schaerer}(2003)}]{schaerer2003}
{Schaerer}, D. 2003, \aap, 397, 527

\bibitem[{{Shirazi} \& {Brinchmann}(2012)}]{Shirazi2012Strongly-star-f}
{Shirazi}, M. \& {Brinchmann}, J. 2012, \mnras, 421, 1043

\bibitem[{Stanway \& Eldridge(2018)}]{Stanway2018Initial-Mass-Fu}
Stanway, E.~R. \& Eldridge, J.~J. 2018, arXiv e-prints, 1811.03856

\bibitem[{{Stark}(2016)}]{Stark2016Galaxies-in-the}
{Stark}, D.~P. 2016, \araa, 54, 761

\bibitem[{{Stasi{\'n}ska} {et~al.}(2015){Stasi{\'n}ska}, {Izotov}, {Morisset},
  \& {Guseva}}]{Stasinska2015Excitation-prop}
{Stasi{\'n}ska}, G., {Izotov}, Y., {Morisset}, C., \& {Guseva}, N. 2015, \aap,
  576, A83

\bibitem[{{Stasi{\'n}ska} \& {Tylenda}(1986)}]{Stasinska1986Intermediate-ma}
{Stasi{\'n}ska}, G. \& {Tylenda}, R. 1986, \aap, 155, 137

\bibitem[{{Steidel} {et~al.}(2016){Steidel}, {Strom}, {Pettini}, {Rudie},
  {Reddy}, \& {Trainor}}]{Steidel2016Reconciling-the}
{Steidel}, C.~C., {Strom}, A.~L., {Pettini}, M., {et~al.} 2016, \apj, 826, 159

\bibitem[{{Sz{\'e}csi} {et~al.}(2015){Sz{\'e}csi}, {Langer}, {Yoon}, {Sanyal},
  {de Mink}, {Evans}, \& {Dermine}}]{Szecsi2015Low-metallicity}
{Sz{\'e}csi}, D., {Langer}, N., {Yoon}, S.-C., {et~al.} 2015, \aap, 581, A15

\bibitem[{{Thuan} {et~al.}(2004){Thuan}, {Bauer}, {Papaderos}, \&
  {Izotov}}]{Thuan2004Chandra-Observa}
{Thuan}, T.~X., {Bauer}, F.~E., {Papaderos}, P., \& {Izotov}, Y.~I. 2004, \apj,
  606, 213

\bibitem[{{Thuan} \& {Izotov}(2005)}]{Thuan2005High-Ionization}
{Thuan}, T.~X. \& {Izotov}, Y.~I. 2005, \apjs, 161, 240

\bibitem[{{Tremmel} {et~al.}(2013){Tremmel}, {Fragos}, {Lehmer}, {Tzanavaris},
  {Belczynski}, {Kalogera}, {Basu-Zych}, {Farr}, {Hornschemeier}, {Jenkins},
  {Ptak}, \& {Zezas}}]{Tremmel2013Modeling-the-Re}
{Tremmel}, M., {Fragos}, T., {Lehmer}, B.~D., {et~al.} 2013, \apj, 766, 19

\bibitem[{{Tumlinson} \& {Shull}(2000)}]{Tumlinson2000Zero-Metallicit}
{Tumlinson}, J. \& {Shull}, J.~M. 2000, \apjl, 528, L65

\bibitem[{{Tzanavaris} {et~al.}(2013){Tzanavaris}, {Fragos}, {Tremmel},
  {Jenkins}, {Zezas}, {Lehmer}, {Hornschemeier}, {Kalogera}, {Ptak}, \&
  {Basu-Zych}}]{Tzanavaris2013Modeling-X-Ray-}
{Tzanavaris}, P., {Fragos}, T., {Tremmel}, M., {et~al.} 2013, \apj, 774, 136

\bibitem[{{van Bever} \& {Vanbeveren}(1998)}]{van-Bever1998The-rejuvenatio}
{van Bever}, J. \& {Vanbeveren}, D. 1998, \aap, 334, 21

\bibitem[{{Vanzella} {et~al.}(2016){Vanzella}, {De Barros}, {Cupani}, {Karman},
  {Gronke}, {Balestra}, {Coe}, {Mignoli}, {Brusa}, {Calura}, {Caminha},
  {Caputi}, {Castellano}, {Christensen}, {Comastri}, {Cristiani}, {Dijkstra},
  {Fontana}, {Giallongo}, {Giavalisco}, {Gilli}, {Grazian}, {Grillo},
  {Koekemoer}, {Meneghetti}, {Nonino}, {Pentericci}, {Rosati}, {Schaerer},
  {Verhamme}, {Vignali}, \& {Zamorani}}]{Vanzella2016High-resolution}
{Vanzella}, E., {De Barros}, S., {Cupani}, G., {et~al.} 2016, \apjl, 821, L27

\bibitem[{{Wiktorowicz} {et~al.}(2018){Wiktorowicz}, {Lasota}, {Middleton}, \&
  {Belczynski}}]{Wiktorowicz2018The-observed-vs}
{Wiktorowicz}, G., {Lasota}, J.-P., {Middleton}, M., \& {Belczynski}, K. 2018,
  arXiv e-prints, 1811.08998

\bibitem[{{Xiao} {et~al.}(2018){Xiao}, {Stanway}, \&
  {Eldridge}}]{Xiao2018Emission-line-d}
{Xiao}, L., {Stanway}, E.~R., \& {Eldridge}, J.~J. 2018, \mnras, 477, 904

\end{thebibliography}

\end{document}